\documentclass{raa}

\usepackage[varg]{txfonts}

\usepackage{natbib}
\usepackage{graphicx}
\usepackage{epstopdf}
\usepackage{amssymb}
\usepackage{float}

\bibpunct{(}{)}{;}{a}{}{,}

\begin{document}

\title{Physical Properties and Catalogue of EW-type Eclipsing Binaries Observed by LAMOST}

   \volnopage{Vol.0 (200x) No.0, 000--000}      
   \setcounter{page}{1}          

\author{Qian S.-B.
      \inst{1,2,3,4}
   \and He J.-J.
      \inst{1,2,3}
   \and Zhang J.
      \inst{1,2,3}
   \and Zhu L.-Y.
      \inst{1,2,3,4}
   \and Shi X.-D.
      \inst{1,2,3}
  \and Zhao E.-G.
      \inst{1,2,3,4}
\and Zhou X.
      \inst{1,2,3}
}

\institute{
             Yunnan Observatories, Chinese Academy of Sciences (CAS), P.O. Box 110, 650011 Kunming, P. R. China {\it qsb@ynao.ac.cn}\\
        \and
             Key Laboratory of the Structure and Evolution of Celestial Objects, Chinese Academy of Sciences, P. O. Box 110, 650216 Kunming, China\\
        \and
             Center for Astronomical Mega-Science, Chinese Academy of Sciences, 20A Datun Road, Chaoyang Dis-trict, Beijing, 100012, P. R. China\\
        \and
             University of the Chinese Academy of Sciences, Yuquan Road 19\#, Sijingshang Block, 100049 Beijing, P. R. China\\
}

\date{Received~~2017 month day; accepted~~2017~~month day}

\def\gsim{\mathrel{\raise.5ex\hbox{$>$}\mkern-14mu
                \lower0.6ex\hbox{$\sim$}}}

\def\lsim{\mathrel{\raise.3ex\hbox{$<$}\mkern-14mu
               \lower0.6ex\hbox{$\sim$}}}

\abstract{
EW-type eclipsing binaries (hereafter EWs) are strong interacting systems where both component stars are usually filling the critical Roche lobes and are sharing a common envelope. Numerous EWs were discovered by several deep photometric survey and there are about 40785 EW-type binary systems listed in the international variable star index (VSX) by March 13, 2017. 7938 of them were observed by LAMOST by November 30, 2016 and their spectral types were given. Stellar atmospheric parameters of 5363 EW-type binary stars were determined based on good spectroscopic observations. In the paper, those EWs were catalogued and their properties are analyzed. The distributions of the orbital period (P), the effect temperature (T), the gravitational acceleration (Log(g)), the metallicity ([Fe/H]) and the radial velocity (RV) are presented for those observed EW-type systems. It is shown that about 80.6\% sample stars have metallicity below zero indicating that EW-type systems are old stellar population. This is in agreement with the the conclusion that the EW binaries are formed from moderately close binaries through angular momentum loss via magnetic braking that takes a few hundred million to a few billion years. The unusual high metallicities of a few percent of EWs may be caused by contaminating of material from the evolution of unseen neutron stars and black holes in the systems. The correlations between the orbital period and the effect temperature, the gravitational acceleration and the metallicity are presented and their scatters are mainly caused by (i) the presence of the third bodies and (ii) the wrong determined periods sometimes. It is shown that some EW contain evolved component stars and the physical properties of EWs are mainly depending on their orbital periods. It is found that the extremely short-period EWs may be older than their long-period cousins because they have lower metallicities. This reveals that they have a longer timescale of pre-contact evolution and their formation and evolution are mainly driven by angular momentum loss via magnetic braking.
}

\authorrunning{Qian et al.}
\titlerunning{EWs Observed by LAMOST}

\maketitle


\section{Introduction}\label{intro}

\begin{figure}
\begin{center}
\includegraphics[angle=0,scale=0.5]{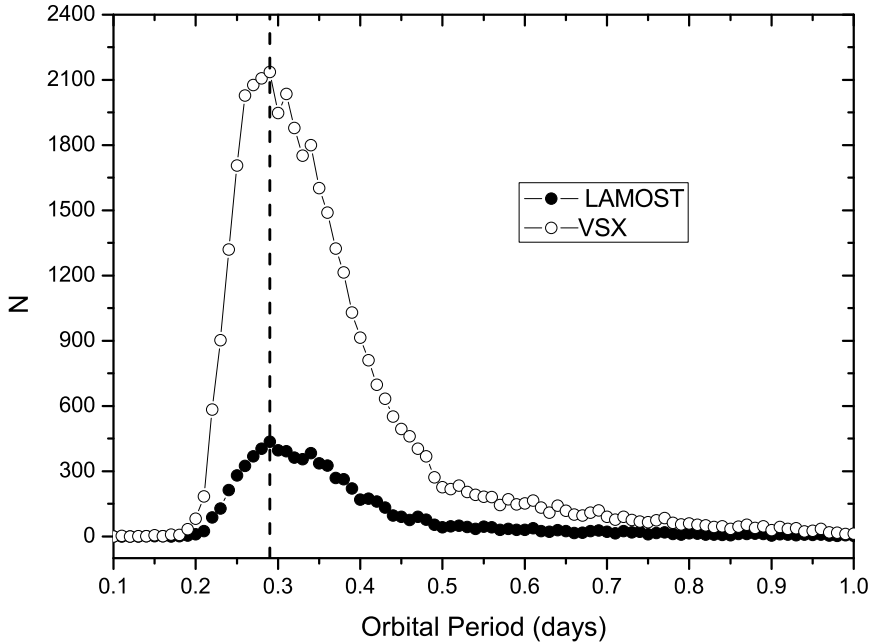}
\caption{Period distribution of EWs. Open circles refers to the systems listed in VSX catalogue, while solid dots to those binaries observed by LAMOST. The peaks of the two period distributions are near 0.29 days (the dashed line).}
\end{center}
\end{figure}

EWs usually consist of two ellipsoidal FGK dwarfs that are in contact with each other and sharing a common convective envelope (CCE) that is lying between the inner and outer critical Roche-lobe surfaces. They are W Ursae Majoris-type eclipsing variables with periods shorter than one days. The light variation is continuous and it is impossible to specify the exact times of onset and end of eclipses. Their light amplitudes are usually $<0.8$\,mag in V and the depths of the primary and secondary minima are almost equal \citep[e.g.,][]{2017ARep...61...80S}. This indicates that the two components possess almost identical temperature and are in thermal contact in spite of different component masses. This kind of stars are different from EB-type binaries whose depths of the primary and secondary minima are not equal and they are not in thermal contact. EW-type binaries are usually detected in older open clusters and globular clusters \citep[e.g.,][]{1993BS....345G}, while they are absent in young seller clusters \citep[e.g.,][]{1998AJ....116.2998R}. Based on these observational facts, some investigators assumed that the EW-type binaries form from short-period detached binaries through angular momentum loss via magnetic braking \citep[e.g.,][]{1988FE....345G, 1994ASPC...56..228B}. The timescale of pre-contact evolution is from a few hundred million to a few billion years. However, how do EW binaries form is still unknown. It is possible that third bodies may play an important role for the origin of the EWs by removing angular momentum from the central binary through early dynamical interaction and/or later evolution \citep[e.g.,][]{2014ApJS..212....4Q, 2013AJ....146...28Z}.

Thanks to several photometric surveys in the world, such as Catalina Sky Survey\footnote{http://www.lpl.arizona.edu/css/}  \cite[CSS,][]{2009ApJ...696..870D,2014ApJS..213....9D}, the asteroid survey LINEAR\footnote{https://astroweb.lanl.gov/lineardb/} \citep{2013AJ....146..101P}, All Sky Automated Survey\footnote{http://www.astrouw.edu.pl/asas/} \citep[ASAS,][]{1997AcA....47..467P, 2005AcA....55..275P} and Northern sky variability survey\footnote{http://www.skydot.lanl.gov/nsvs/nsvs.php} \citep[NSVS,][]{2004AJ....127.2436W}, a large number of EW binaries were discovered. Since the data on variable stars including EWs are constantly changing, the mission of VSX\footnote{http://www.aavso.org/vsx/} \citep[the international variable star index, ][]{2006SASS...25...47W} is to bring all of that new information together in a single data repository and provides the tools necessary for the controlled and secure revising of the data. 40785 EW-type binary systems were listed in VSX by March 13, 2017. Among the 40785 EWs, the orbital periods of 40646 systems were given. Those survey data are very useful to understand the photometric properties of EW binaries. However, statistically spectroscopic properties of those sample stars are unclear because of the lack of spectral surveys. The light curves of many EW binaries were solved recently, but their spectral types are usually unknown.

The Large Sky Area Multiobject Fiber Spectroscopic Telescope (LAMOST, also called as Guo Shou Jing telescope) is a special telescope with an effective aperture about 4 meters that located at Xinglong station, National Astronomical Observatories of China (NAOC). It has a field of view of 5 degrees and could simultaneously obtain the spectra of about 4000 stars with low-resolution of about 1800 in one exposure \citep{1996ApOpt..35.5155W, 2012RAA....12.1197C}. The wavelength range of LAMOST is from 3700 to 9000\,{\AA} and is divided in two arms, i.e., a blue arm (3700-5900\,{\AA}) and a red arm (5700-9000\,{\AA}). The final spectrum of each target is obtained by merging several exposures. Huge amounts of spectroscopic data have been obtained \citep[e.g.,][]{2012RAA....12..723Z, 2012RAA....12.1243L, 2015RAA....15.1095L}.

In the recent LAMOST data release, about 19.5\% EW-type binaries (7938) in VSX were observed by LAMOST survey from October 24, 2011 to November 30, 2016. Among the 7938 EWs, the orbital periods of 7930 samples are given in VSX. The distribution of the orbital period for those observed EWs by LAMOST is shown in Fig. 1. Also displayed in the figure is the period distribution of all EWs in VSX where 139 EWs without orbital periods are not shown. Those spectroscopic data can be used during the photometric solutions and the big data of stellar spectrum from LAMOST survey provide important information for studying EWs. In the paper, EW binaries observed in the LAMOST survey are catalogued. Then, based on the distributions of those atmospheric parameters and some statistical correlations, the physical properties and the formation and evolutionary states of EW binaries are discussed.

\section{Catalogue of EWs observed by LAMOST}

In the recent LAMOST data release, about 7938 EWs in the VSX catalogue were observed from October 24, 2011 to November 30, 2016 and their spectral types were obtained. Among the 7938 EWs, the stellar atmospheric parameters of 5363 systems were determined when their spectra have higher signal to noise. The stellar atmospheric parameters including the effect temperature $T_{eff}$, the gravitational acceleration Log (g), the metallicity [Fe/H] and the radial velocity $V_{r}$ were automatically derived by the LAMOST stellar parameter pipeline when their spectra are good and reliable \citep{2011RAA....11..924W, 2014IAUS..306..340W, 2015RAA....15.1095L}. Those stellar atmospheric parameters were determined based on the Universite de Lyon spectroscopic analysis software (ULySS) \citep{2009A&A...501.1269K, 2011A&A...525A..71W}. The ULySS fits the full observed spectra by using the model spectrum that is generated by an interpolator by using the ELODIE library as a reference \citep[e.g., ][]{2001A&A...369.1048P, 2007astro.ph..3658P}. When $T_{eff} < 8000$\,K, the standard deviations are 110\,K, 0.19\,dex and 0.11\,dex for $T_{eff}$, Log (g)g and [Fe/H] respectively. For the radial velocity $V_{r}$, the standard deviations are 4.91\,Km/s when $T_{eff} < 10000$\,K \citep[e.g.,][]{2015RAA....15.2204G}.

\begin{figure}
\begin{center}
\includegraphics[angle=0,scale=0.5]{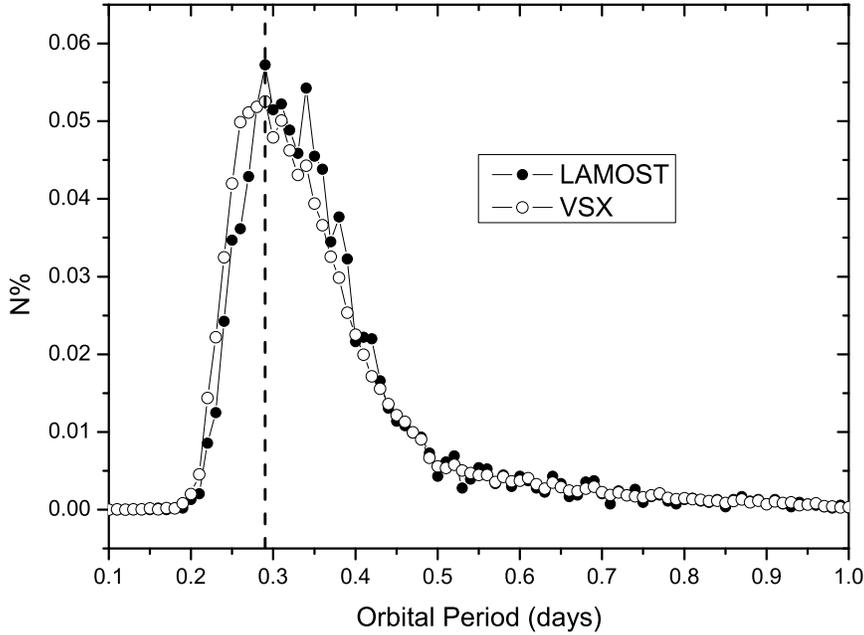}
\caption{Relative distribution of orbital period for EWs. Open circles refers to all EWs listed in VSX catalogue, while solid dots to those 5363 EWs whose stellar atmospheric parameters were determined by using LAMOST data. The period distribution peaks are near 0.29 days (the dashed line).}
\end{center}
\end{figure}

The observations of the 5363 EWs are catalogued in the order of increasing VSX number. Some EWs were observed trice or more times on different dates and we list all of the parameters. Those listed in Table 1 are the first 20 observations. The whole catalogue is available through the internet (the electronic version of the catalogue is at the website\footnote{http://search.vbscn.com/CEW.table1.txt}) and it will be improved by adding new data obtained by LAMOST in the future. The table include binary names, their right ascensions (RA) and declinations (DEC), types of light variation and orbital periods. These parameters are from VSX catalogue. Those shown in column 6 are the distances (in arcsec) between the two positions determined by the coordinates given in SVX and by LAMOST. The distances were used to identify those EWs from the LAMOST samples based on the criterion Dist$<2$\,arcsecs. The observing dates are listed in column 7, while the determined spectral types of those EWs are shown in column 8. The stellar atmospheric parameters, $T_{eff}$, Log (g), [Fe/H] and $V_{r}$) of the 5363 EWs, are listed in columns 9, 11, 13 and 15. $E_1$, $E_2$, $E_3$ and $E_4$ in the table are their errors respectively.

\begin{table*}
\center
\tiny
\caption{Catalogue of EWs observed by LAMOST (the first 20 observations).}\label{XXXX}
\begin{tabular}{lllllllllllllllll}\hline\hline
Name       &  R.A.     & Dec.       &Type    &  P (days)  & Dist   & Date       & Sp.  & T (K)   & $E_1$  & Log(g) & $E_2$ &[Fe/H]  & $E_3$  & RV      & $E_4$ \\\hline
HU And     & 003929.69 & +400459.8  &EW      & 0.285789   & 0.360  & 2011-10-28 &  G9  & 5158.97 & 246.09 & 4.431  & 0.352 &  0.126 & 0.229  &    0.73 & 18.92 \\
IM And     & 004646.80 & +393833.0  &EW      & 0.270377   & 1.108  & 2014-12-12 &  K3  & 4764.13 & 208.42 & 4.182  & 0.299 & -0.143 & 0.194  & -136.85 & 15.35 \\
IM And     & 004646.80 & +393833.0  &EW      & 0.270377   & 1.108  & 2015-09-13 &  K5  & 4977.45 &  73.29 & 4.692  & 0.105 & -0.029 & 0.068  & -157.15 &  5.28 \\
LY And     & 022152.54 & +383742.4  &EW      & 0.34505    & 0.613  & 2013-11-12 &  G2  & 5751.81 &  29.16 & 4.090  & 0.041 & -0.315 & 0.027  &    0.16 &  2.75 \\
MT And     & 022704.78 & +400328.5  &EW      & 0.358781   & 0.767  & 2013-11-12 &  G3  & 5676.33 & 204.28 & 3.931  & 0.293 & -0.136 & 0.190  &   14.41 & 15.32 \\
MT And     & 022704.78 & +400328.5  &EW      & 0.358781   & 0.911  & 2013-10-14 &  G3  & 5751.14 &  25.60 & 4.144  & 0.035 &  0.005 & 0.024  &  -17.61 &  2.62 \\
QX And     & 015757.78 & +374822.5  &EW      & 0.4121753  & 0.045  & 2014-11-10 &  F5  & 6501.66 &   2.24 & 4.174  & 0.001 & -0.066 & 0.002  &    4.53 &  0.49 \\
GK Aqr     & 221956.93 & -007986.9  &EW      & 0.3274145  & 0.344  & 2016-11-03 &  K0  & 5302.35 &   7.27 & 4.274  & 0.005 &  0.384 & 0.006  &  -17.35 &  1.81 \\
GK Aqr     & 221956.93 & -007986.9  &EW      & 0.3274145  & 1.188  & 2012-10-04 &  K1  & 5365.27 &  56.54 & 4.368  & 0.080 &  0.458 & 0.053  &  -55.10 &  5.00 \\
GM Aqr     & 222157.94 & -028042.8  &EW      & 0.3672853  & 0.491  & 2016-11-03 &  G7  & 5636.36 &  21.72 & 4.296  & 0.030 &  0.153 & 0.021  &  -66.66 &  2.33 \\
GM Aqr     & 222157.94 & -028042.8  &EW      & 0.3672853  & 0.491  & 2012-10-04 &  G7  & 5715.99 & 132.99 & 4.308  & 0.190 &  0.109 & 0.124  &   34.44 & 10.36 \\
GS Aqr     & 222733.63 & -005757.6  &EW      & 0.374067   & 1.671  & 2015-11-09 & A5V  & 7023.10 &   3.25 & 4.276  & 0.002 & -1.038 & 0.003  &  -53.16 &  0.81 \\
GS Aqr     & 222733.63 & -005757.6  &EW      & 0.374067   & 0.479  & 2016-11-03 & A7V  & 7027.32 &   7.89 & 4.323  & 0.010 & -0.938 & 0.008  & -112.34 &  1.25 \\
AH Aur     & 062604.93 & +275956.4  &EW/KW   & 0.494106   & 0.267  & 2012-02-01 &  G3  & 6017.81 &  14.79 & 4.025  & 0.020 &  0.399 & 0.014  &   48.62 &  1.68 \\
V0468 Aur  & 045611.66 & +444638.9  &EW      & 0.91278951 & 0.050  & 2012-02-09 &  F5  & 6548.18 & 182.53 & 3.856  & 0.262 & -0.001 & 0.170  &   13.64 & 13.48 \\
TU Boo     & 140458.04 & +300001.5  &EW/KW   & 0.3242868  & 0.144  & 2014-03-25 &  G3  & 5828.78 &   9.41 & 4.295  & 0.012 & -0.027 & 0.009  &    7.13 &  1.32 \\
TU Boo     & 140458.04 & +300001.5  &EW/KW   & 0.3242868  & 0.144  & 2016-05-18 &  G3  & 5710.93 &   9.32 & 4.165  & 0.012 & -0.095 & 0.009  &  -13.39 &  1.24 \\
TZ Boo     & 150809.13 & +395812.9  &EW/KW   & 0.297162   & 0.081  & 2016-02-25 &  G2  & 5597.28 &   6.21 & 4.179  & 0.008 & -0.746 & 0.006  &  -59.65 &  0.92 \\
AK Boo     & 133839.06 & +241105.4  &EW/KE   & 0.694030   & 0.436  & 2015-03-19 &  G7  & 5526.82 &  95.45 & 4.065  & 0.136 & -0.326 & 0.089  &   45.21 &  7.77 \\
AQ Boo     & 134726.86 & +171824.7  &EW      & 0.333139   & 0.936  & 2014-03-06 &  G0  & 5680.80 &   2.64 & 4.037  & 0.002 & -0.466 & 0.002  &  -52.13 &  0.59 \\
\hline\hline
\end{tabular}
\end{table*}

The temperatures of most EWs are lower than 8000\,K and their standard deviations could be estimated reliably. However, both components of EWs are rapidly rotating and highly deformed stars that share a common envelope. Do their spectra have sufficient tracers necessary for unique determination of atmospheric parameters? There are 25 EWs were observed five times or more. To check the reliability of stellar atmospheric parameters and to answer the question, we determined the mean values of their atmospheric parameters and derived the corresponding standard errors. The results are shown in Table 2 where their names and orbital periods are listed in the first and the second columns. The observational times are shown in third column, while the average atmospheric parameters and their standard errors are displayed in the rest columns. As shown in Table 2, the standard errors of the effect temperature for all targets are lower than 110\,K. Apart from two targets, the standard errors of the gravitational acceleration Log (g) for the rest targets are lower than 0.19\,dex. The standard errors of the metallicity for most EWs targets are lower than 0.11\,dex. These results may indicate that it is no matter that we could use single star spectra to calibrate peanut-shaped stars and demonstrate that the extracting parameters could reach the mentioned level of precision. Moreover, those EWs were observed at different phases, those results also reveal that there are no obvious effects of phase on the derived atmospheric parameters across multiple observations of the same object within the errors.

\begin{table}[!h]
\scriptsize
\caption{Mean atmospheric parameters for 25 EWs observed more than 4 times and their standard errors.}\label{table1}
\begin{center}
\begin{tabular}{llccccccc}\hline
 Star Name              &  P (days)  &  Times  &   $\overline{T_{eff}}$(K) &   Errors      &   $\overline{Log(g)}$   &  Errors   &   $\overline{[Fe/H]}$ & Errors  \\\hline
T-Lyr1-15705            &  0.298933  &   8     &     5473.44               &    68.40      &     4.340               &   0.101   &      0.117            &   0.070 \\
LINEAR 6102187          &  0.324026  &   7     &     5479.31               &    94.55      &     4.234               &   0.153   &     -0.047            &   0.041 \\
CSS J074328.7+360725    &  0.391234  &   6     &     6158.67               &    67.58      &     4.157               &   0.072   &     -0.084            &   0.123 \\
CSS J222851.7+065242    &  0.370323  &   5     &     6687.61               &   109.41      &     4.064               &   0.104   &     -0.322            &   0.089 \\
CSS J111218.6+542629    &  0.317854  &   5     &     5421.22               &    61.99      &     4.339               &   0.072   &     -0.042            &   0.031 \\
CSS J093510.3+313745    &  0.295414  &   5     &     6056.15               &    72.54      &     4.089               &   0.090   &     -0.572            &   0.017 \\
CSS J090411.9+132908    &  0.273486  &   5     &     5135.15               &    33.33      &     4.165               &   0.090   &     -0.404            &   0.058 \\
CSS J082217.4+064452    &  0.340438  &   5     &     5780.39               &    89.51      &     4.070               &   0.192   &     -0.236            &   0.093 \\
CSS J073220.4+283612    &  0.274257  &   5     &     5330.87               &    47.72      &     4.332               &   0.113   &     -0.189            &   0.041 \\
CSS J070315.8+423814    &  0.368082  &   5     &     5767.00               &    62.65      &     4.173               &   0.080   &      0.099            &   0.041 \\
CSS J025901.9+313046    &  1.0544407 &   5     &     6304.31               &    46.60      &     3.754               &   0.091   &      0.148            &   0.068 \\
CSS J025613.4+311517    &  0.538159  &   5     &     6237.42               &    46.37      &     4.106               &   0.020   &     -0.083            &   0.034 \\
CSS J022913.6+042841    &  0.29825   &   5     &     5838.94               &    48.30      &     4.187               &   0.166   &     -0.755            &   0.189 \\
CSS J011822.4+080543    &  0.454696  &   5     &     6119.61               &    56.82      &     4.131               &   0.124   &     -0.557            &   0.017 \\
CSS J011310.1+370333    &  0.39251   &   5     &     6077.21               &    65.25      &     4.123               &   0.055   &     -0.177            &   0.051 \\
LINEAR 6446092          &  0.30679788&   5     &     5484.59               &    43.30      &     4.381               &   0.196   &      0.172            &   0.068 \\
NSVS 4831297            &  0.369811  &   5     &     5926.45               &    24.09      &     4.202               &   0.086   &      0.202            &   0.053 \\
KID 06964796            &  0.399961  &   5     &     5939.73               &    27.60      &     4.237               &   0.044   &      0.133            &   0.014 \\
KID 11084782            &  0.586555  &   5     &     8222.08               &    17.16      &     4.011               &   0.018   &     -0.269            &   0.025 \\
NSVS 4583537            &  0.40835410&   5     &     6284.33               &    66.72      &     4.120               &   0.042   &     -0.111            &   0.049 \\
VSX J065147.6+592649    &  0.4217    &   5     &     5996.10               &    35.18      &     4.127               &   0.079   &      0.411            &   0.018 \\
NY Boo                  &  0.32679   &   5     &     5779.52               &    30.98      &     4.115               &   0.055   &     -0.076            &   0.027 \\
NSVS 7209962            &  0.295931  &   5     &     5276.37               &    37.39      &     4.273               &   0.080   &      0.348            &   0.055 \\
OP Leo                  &  0.391931  &   5     &     6104.86               &    51.49      &     4.129               &   0.061   &     -0.385            &   0.053 \\
ASAS J081149-0111.1     &  0.567286  &   5     &     6275.02               &    93.33      &     4.346               &   0.030   &      0.143            &   0.034 \\
V0449 Gem               &  0.270659  &   5     &     5045.36               &    42.53      &     4.321               &   0.049   &     -0.048            &   0.026 \\
V1022 Tau               &  0.34717194&   5     &     5882.69               &    52.00      &     4.193               &   0.049   &     -0.255            &   0.045 \\
\hline
\end{tabular}
\end{center}
\end{table}

The relative distribution (the percentage of the number to the whole sample) of the orbital period for the 5363 EWs is displayed in Fig. 2. For comparison, the relative period distribution of all EWs in VSX is also shown in the figure. It is found that both of them are overlapping nearly. This indicates that the 5363 EWs could be used to represent the properties of the whole EWs in the total VSX catalogue. As shown in Figs. 1 and 2, the period distribution peaks are near 0.29\,days (the dashed line). This is shorter than that given by \cite{2006MNRAS.368.1311P} who obtain a peak near 0.37\,days based on the ASAS data \citep[e.g.,][]{1997AcA....47..467P, 2005AcA....55..275P}. This may be caused by the fact that ASAS is dedicated to the detection of the variability of bright stars, while many faint short-period EWs were discovered by recent deep photometric surveys \citep[e.g.,][]{2009ApJ...696..870D, 2014ApJS..213....9D}. As those detected by several investigators \citep[e.g.,][]{2006MNRAS.368.1311P,  2011ApJ...731...17B}, it has a sharp cut-off at 0.2\,days. A long tail is extending beyond 1\,d and the tail actually extends to about 24\,days.

For some LAMOST spectra of EW-type binaries, their signals to noise are not high enough to determine the stellar atmospheric parameters. In the case, only spectral types were given. The spectral types of those EWs are also catalogued in the order of increasing VSX number. Those shown in Table 3 are the first 20 spectral types in the catalogue. The whole table is available at the website\footnote{http://search.vbscn.com/CEW.table3.txt} via the internet. The catalogue lists 3732 spectral types for 3055 EWs. The descriptions of those columns are the same as those in Table 1. For about 1691 EWs, only spectral type was obtained by LAMOST. Both spectral types and stellar atmospheric parameters were determined for 5363 EWs. For the rest ones, no results were obtained.

\begin{table*}[h]
\footnotesize
\begin{center}
\caption{Spectral types determined by LAMOST (the first 20 observations).}\label{XXXX}
\begin{tabular}{llllllll}\hline\hline
Name      & R.A.       & Dec.     & Type      &  Period (days) & Distance & Date        & Sp.   \\\hline
MS And    &  022546.86 &+395845.5 & EW        &  0.777900      & 0.559    & 2013-11-12  &    K3 \\
GS Aqr    &  222733.63 &-005757.6 & EW        &  0.374067      & 0.479    & 2012-10-29  &   A5V \\
MR Aur    &  055133.68 &+310652.0 & EW        &  0.690301      & 1.541    & 2011-11-10  &  A2IV \\
MR Aur    &  055133.68 &+310652.0 & EW        &  0.690301      & 1.541    & 2011-12-25  &   A1V \\
CK Boo    &  143503.76 &+090649.4 & EW/RS     &  0.355152      & 0.093    & 2016-04-24  &    F0 \\
GH Boo    &  141451.51 &+273415.7 & EW        &  0.65951       & 0.833    & 2015-01-15  & A8III \\
GH Boo    &  141451.51 &+273415.7 & EW        &  0.65951       & 0.053    & 2016-05-15  &    G7 \\
GQ Boo    &  145936.67 &+250244.9 & EW        &  0.3846402     & 0.126    & 2012-06-05  &    G8 \\
GW Cnc    &  084812.69 &+210713.8 & EW        &  0.281413      & 0.053    & 2015-03-08  &    G7 \\
UZ CMi    &  075051.76 &+033903.5 & EW/DW     &  0.551361      & 0.180    & 2013-01-23  &    F6 \\
BB CMi    &  075124.55 &+045439.2 & EW        &  0.792866160   & 0.204    & 2013-01-23  &    F0 \\
SS Com    &  124939.08 &+184211.9 & EW/KW     &  0.412822      & 0.076    & 2016-05-19  &    G7 \\
EY Com    &  131355.38 &+310454.1 & EW/KW     &  0.2993278     & 0.715    & 2012-02-26  &    K5 \\
LP Com    &  123305.52 &+270803.6 & EW        &  0.33793358    & 0.107    & 2012-01-23  &    K4 \\
LP Com    &  123305.52 &+270803.6 & EW        &  0.33793358    & 0.107    & 2012-01-11  &    K1 \\
V2213 Cyg &  192857.89 &+430625.5 & EW        &  0.350094      & 0.220    & 2014-09-13  &    G7 \\
V2284 Cyg &  192955.02 &+485500.1 & EW        &  0.306994      & 0.014    & 2013-09-14  &    G7 \\
V2284 Cyg &  192955.02 &+485500.1 & EW        &  0.306994      & 0.004    & 2015-10-01  &    G7 \\
IV Dra    &  153623.26 &+531911.3 & EW        &  0.268105      & 0.017    & 2013-05-03  &    K5 \\
IV Dra    &  153623.26 &+531911.3 & EW        &  0.268105      & 0.201    & 2012-06-17  &    K3 \\
\hline\hline
\end{tabular}
\end{center}
\end{table*}

\section{Distributions of stellar atmospheric parameters for EWs}

\begin{figure}
\begin{center}
\includegraphics[angle=0,scale=0.5]{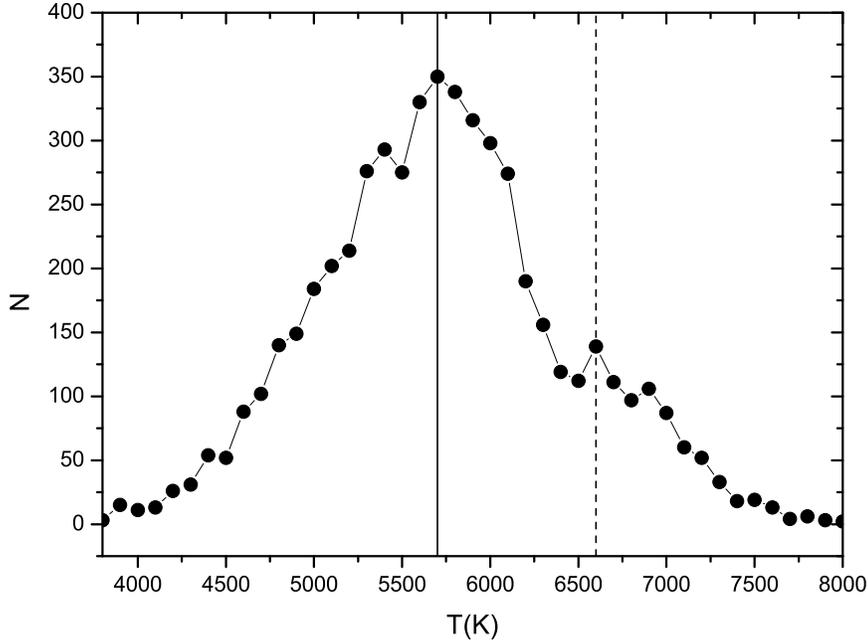}
\caption{Distribution of the effect temperature for EWs observed by LAMOST. The solid and the dashed lines refer to the two peaks near 5700\,K and 6600\,K respectively.}
\end{center}
\end{figure}

\begin{figure}
\begin{center}
\includegraphics[angle=0,scale=0.5]{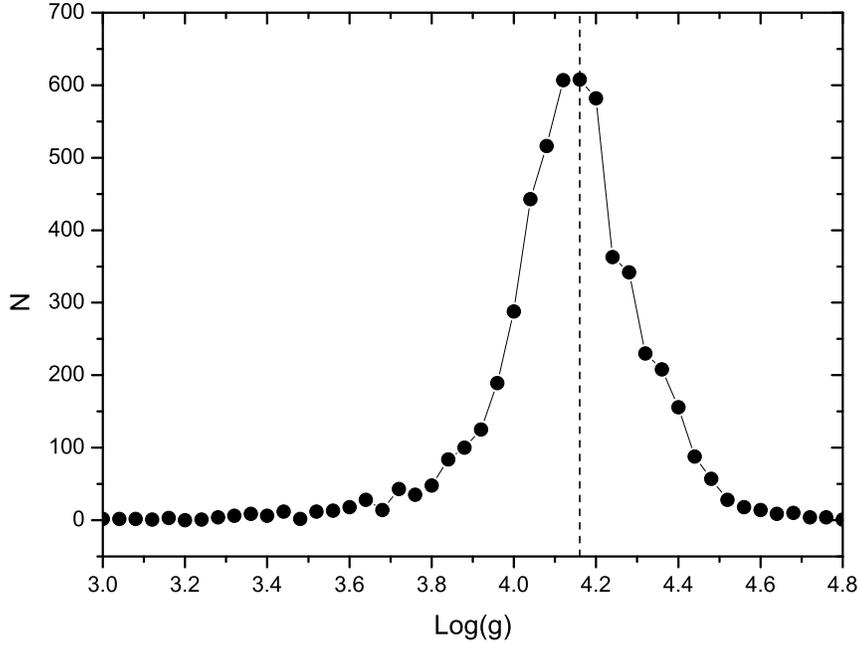}
\caption{Distribution of the gravitational acceleration Log (g) for EWs observed by LAMOST. The dashed line refers to the peak near 4.16.}
\end{center}
\end{figure}

\begin{figure}
\begin{center}
\includegraphics[angle=0,scale=0.5]{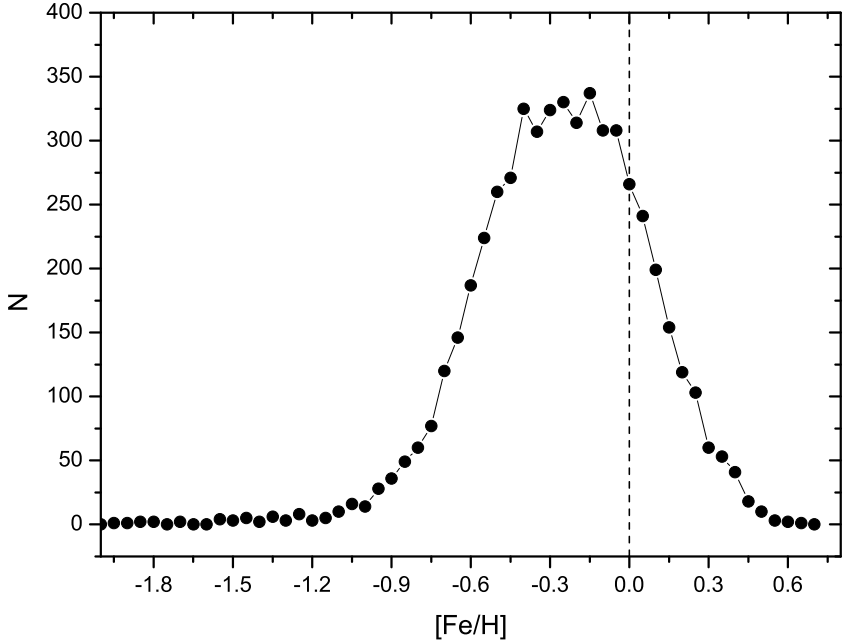}
\caption{Distribution of the metallicity [Fe/H] for EWs observed by LAMOST. It is shown that most of the EWs have $[Fe/H] < 0$.}
\end{center}
\end{figure}

\begin{figure}
\begin{center}
\includegraphics[angle=0,scale=0.5]{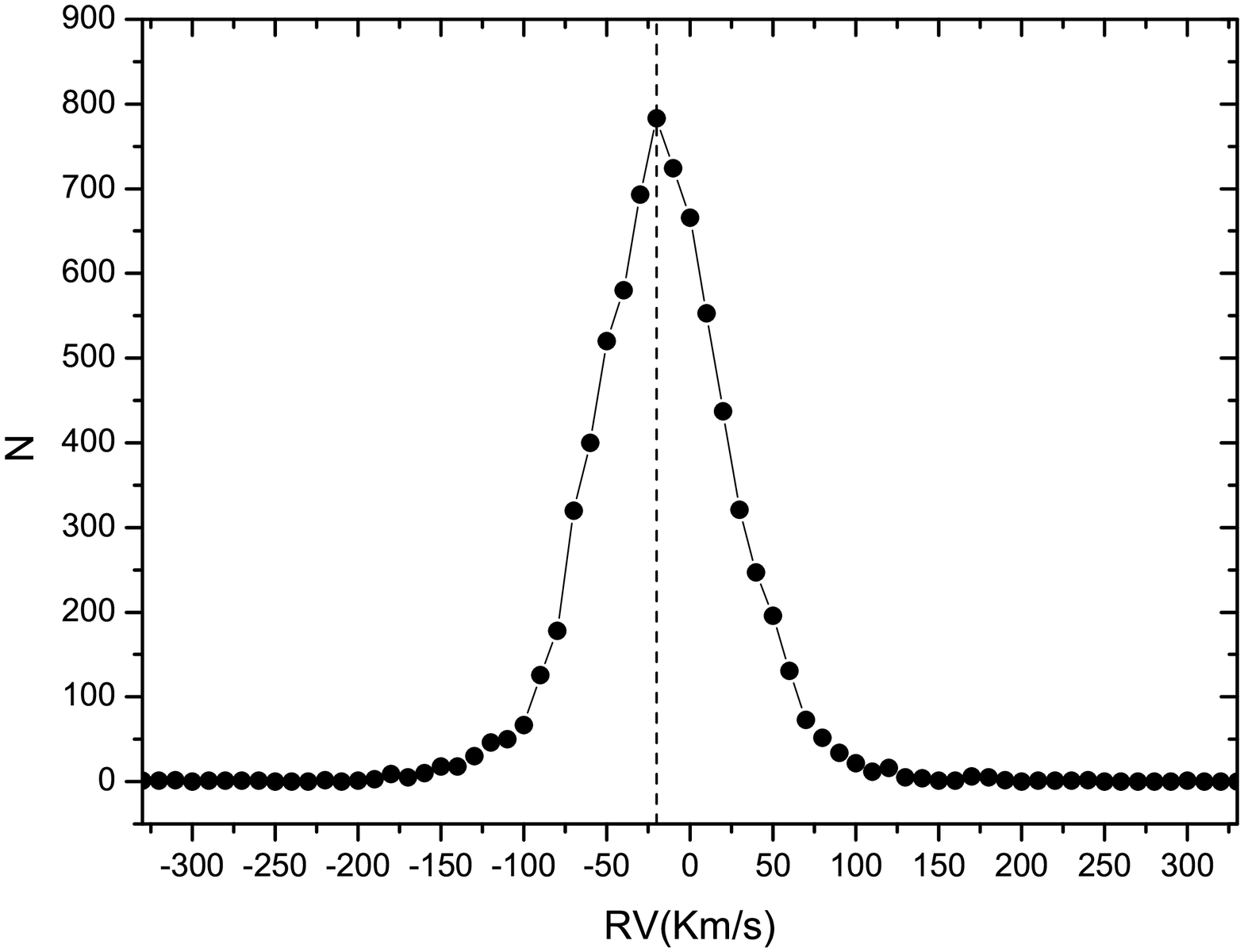}
\caption{Distribution of the radial velocity $V_{r}$ for EWs observed by LAMOST. There is a peak near $V_{r}=-20$\,Km/s.}
\end{center}
\end{figure}

As aforementioned, the stellar atmospheric parameters of 5363 systems were determined and their relative period distribution is the same as that of all EWs in VSX. Therefore, they could be used to investigate the properties of the whole EWs. During the analyses, when the EWs were observed two times or more, the stellar atmospheric parameters, the effect temperature $T_{eff}$, the gravitational acceleration Log (g) and the metallicity [Fe/H] were averaged and we used the mean values. As for the radial velocity $V_{r}$, we did not average them because they were observed at different phases and are varying with time.

The binary temperature distribution is shown in Fig. 3 and the distribution has a main peak near 5700\,K (the solid line). This peak is corresponding to the temperature of a G3-type main-sequence star with a stellar mass about 0.97\,$M_{\odot}$ \citep{2000AAQ...C}. This indicates that majority EWs are solar-type stars and have the proton-proton (p-p) chain nuclear reaction at their center cores. Fig. 3 also shows that there is a small peak near 6600\,K (the dashed line). This corresponds to the temperature of an F6-type main-sequence star with a stellar mass about 1.35\,$M_{\odot}$. From the first peak to the second small peak, it may reflects the transit of center nuclear reaction from the p-p chain to the carbon¨Cnitrogen¨Coxygen (CNO) cycles. The distribution of the gravitational acceleration Log (g) is plotted in Fig. 4. The distribution peaks near 4.16. It is shown that the sample of EW systems are homogeneous and most of EWs are main-sequence binaries. This is in agreement with the idea that EWs are formed from detached main-sequence binaries via the combination of Case A mass transfer and angular momentum loss via magnetic braking \citep[e.g.,][]{2013ApJS..207...22Q}.

The metallicity ([Fe/H]) distribution is shown in Fig. 5. EW-type binaries are usually composed of two solar-type stars. It is expected that their metallicities are similar to those detected among stars in the solar neighborhood \citep[e.g.,][]{2013AJ....146...70R}. However, as shown in Fig. 5, the metallicities of 80.6\% EWs are lower than that of the Sun, i.e., $[Fe/H] < 0$. For stars in the Galaxy, stellar metallicities are weakly correlated with their ages \citep[e.g.,][]{2007ApJ...665..767R, 2009IAUS..258...23F}. The low metallicities indicates that most EWs are old stellar population with longer ages. For a few percent of EWs, their metallicities are higher than 0.3 ($[Fe/H] > 0.3$). The possibility of the unusual high metallicities may be through contamination by material from unseen degenerate objects (e.g., neutron stars and black holes) that orbiting the binaries. Their progenitors are originally much more massive third stars in triple systems. Ten EWs have the highest metallicities are shown in Table 4. They are a good source to search for potential degenerate objects (e.g., neutron stars and black holes) orbiting EWs.

\begin{table}[h]
\footnotesize
\begin{center}
\caption{Ten EWs with the highest metallicities.}\label{XXXX}
\begin{tabular}{lllllll}\hline\hline
Name                     &  Period (days) & Sp.     & T (K)   & Log (g) &[Fe/H]  & RV (Km/s) \\\hline
CSS\_J080956.0+131054    & 0.316903       &     K1  & 5692.04 & 4.491   & 0.677  & -50.76    \\
VSX J001137.3+303145     & 0.41222        &     F9  & 5719.4  & 4.088   & 0.637  & -66.27    \\
CSS\_J083051.6+185801    & 0.331662       &     G5  & 5323.88 & 4.104   & 0.609  & 15.62     \\
NSVS 2729390             & 0.347679       &     K0  & 5490.45 & 4.36    & 0.586  & -63.81    \\
ASAS J163229+0818.1      & 0.399559       &     G8  & 5815.14 & 4.254   & 0.56   & -12.5     \\
NSVS 4231740             & 0.40694612     &  F9/G5  & 5809.96 & 4.085   & 0.557  & 35.28     \\
V1047 Her                & 0.32073733     &     K1  & 5466.2  & 4.211   & 0.545  & -85.15    \\
CSS\_J032658.1+142940    & 0.385214       &     G7  & 5650.16 & 4.336   & 0.524  & -2.23     \\
CSS\_J072417.0+224103    & 0.342338       &  G8/K1  & 5497.9  & 4.233   & 0.519  & -37.3     \\
CSS\_J002629.5+445324    & 0.325968       &  K1/K3  & 5043.48 & 4.398   & 0.514  & -0.99     \\
\hline\hline
\end{tabular}
\end{center}
\end{table}

\begin{table}[h]
\footnotesize
\begin{center}
\caption{EWs observed by LAMOST with radial velocities larger than 200\,Km/s.}\label{XXXX}
\begin{tabular}{lllllll}\hline\hline
Name                          &  Period (days) & Sp.   & T (K)   & Log (g) &[Fe/H]  & RV (Km/s) \\\hline
UY Hya                        & 0.7275         &  A2V  & 6697.05 & 4.233   & -1.834 & 359.43    \\
CSS\_J212703.2+100332         & 0.452204       &  A5V  & 7065.8  & 4.27    & -1.022 & -321.77   \\
CSS\_J212703.2+100332         & 0.452204       &  A3V  & 6817.47 & 4.248   & -1.039 & -316.4    \\
CSS\_J005447.7+284030         & 0.314908       &   F0  & 6737.72 & 4.265   & -0.888 & -306.44   \\
CSS\_J032645.2+004931         & 0.705836       &   F5  & 5826.79 & 4.062   & -1.405 & -304.7    \\
CSS\_J112812.6+045202         & 0.316715       &   F5  & 5985.71 & 3.965   & -0.881 & 301.29    \\
CSS\_J214828.6+090336         & 1.42465        &   F0  & 6611.71 & 4.365   & -1.17  & -285.31   \\
CSS\_J010448.7+373231         & 0.283694       &   F0  & 6139.34 & 4.047   & -1.411 & -276.72   \\
CSS\_J014525.5+360334         & 0.270282       &   G7  & 5342.15 & 4.101   & -0.579 & -264.58   \\
CSS\_J170410.6+274628         & 0.709314       &  A9V  & 6711.52 & 4.136   & -0.627 & -255.78   \\
NSVS 7285749                  & 0.323629       &   F0  & 6438    & 4.322   & -1.018 & 242.96    \\
CSS\_J115356.5-022540         & 0.224256       &   K2  & 5074.82 & 4.745   & -0.917 & 240.42    \\
CSS\_J072136.0+403327         & 0.340065       &   F4  & 6714.6  & 4.409   & 0.467  & 230.7     \\
NSVS 7285749                  & 0.323629       &   F0  & 6477.27 & 4.323   & -0.963 & 222.12    \\
CSS\_J012955.1+391130         & 0.317905       &   F7  & 5932.42 & 4.1     & -0.52  & -216.98   \\
LINEAR 14668373               & 0.224377       &   G8  & 4989.28 & 4.014   & -1.591 & -212.1    \\
CSS\_J032034.5+203607         & 0.320948       &   K5  & 5874.96 & 4.888   & 0.005  & 210.36    \\
\hline\hline
\end{tabular}
\end{center}
\end{table}

The distribution of the radial velocity ($V_{r}$) for those EWs is displayed in Fig. 6. 7382 RVs for 5363 EWs are used for constructing the figure. A peak is near $V_{r}=-20$\,Km/s and the distribution is symmetric. This may reveal that the $V_0$ of most EWs are close to this value. The amplitudes of radial velocity curves for EWs are about 150-300\,Km/s \citep[e.g.,][]{2001AJ....122.1974R}. Fig. 6 reflects a statistical random sampling of radial velocity curves for EWs. Sixteen EWs with radial velocities larger than 200\,Km/s are shown in Table 5. They may be observed near the maxima or the minima of the radial velocity curves of those EWs.

\section{Statistical correlations between the orbital period and the stellar atmosphere parameters}

\begin{figure}
\begin{center}
\includegraphics[angle=0,scale=0.5]{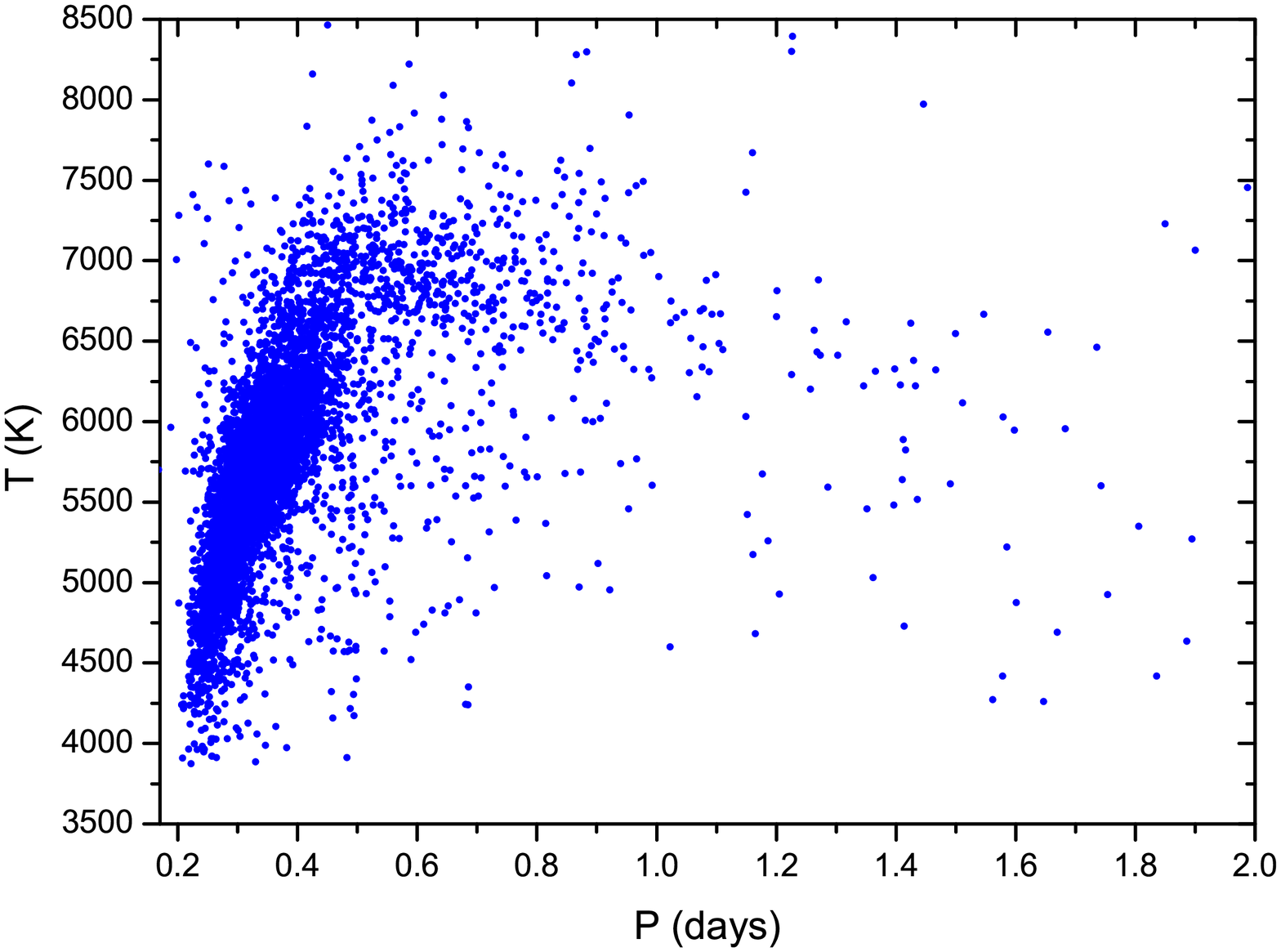}
\caption{The correlation between the orbital period and the effect temperature for EWs observed by LAMOST. 55 EWs with orbital period longer than 2 days are not shown in the figure.}
\end{center}
\end{figure}

\begin{figure}
\begin{center}
\includegraphics[angle=0,scale=0.5]{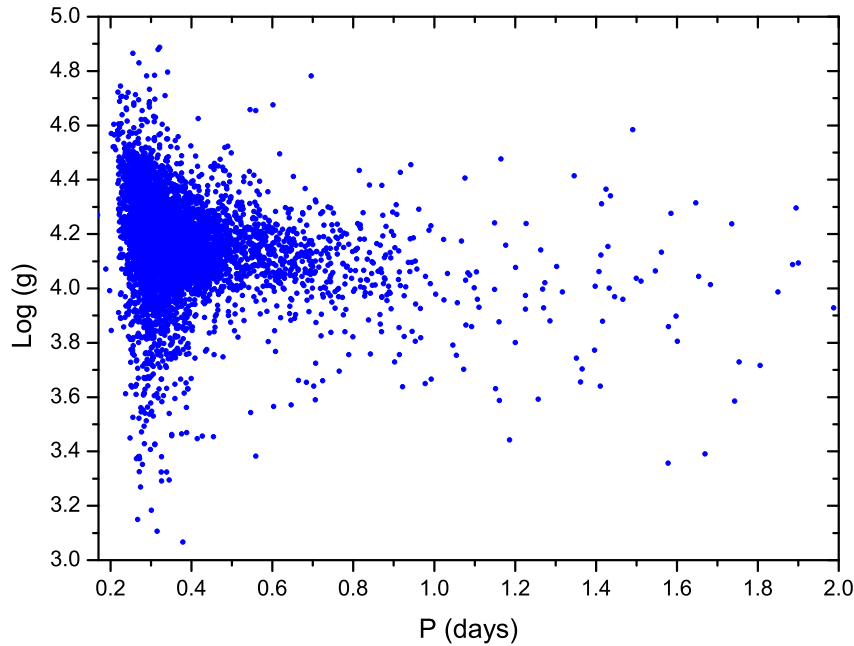}
\caption{The same as those in Fig. 7 but for the correlation between the orbital period and the gravitational acceleration.}
\end{center}
\end{figure}

\begin{figure}
\begin{center}
\includegraphics[angle=0,scale=0.5]{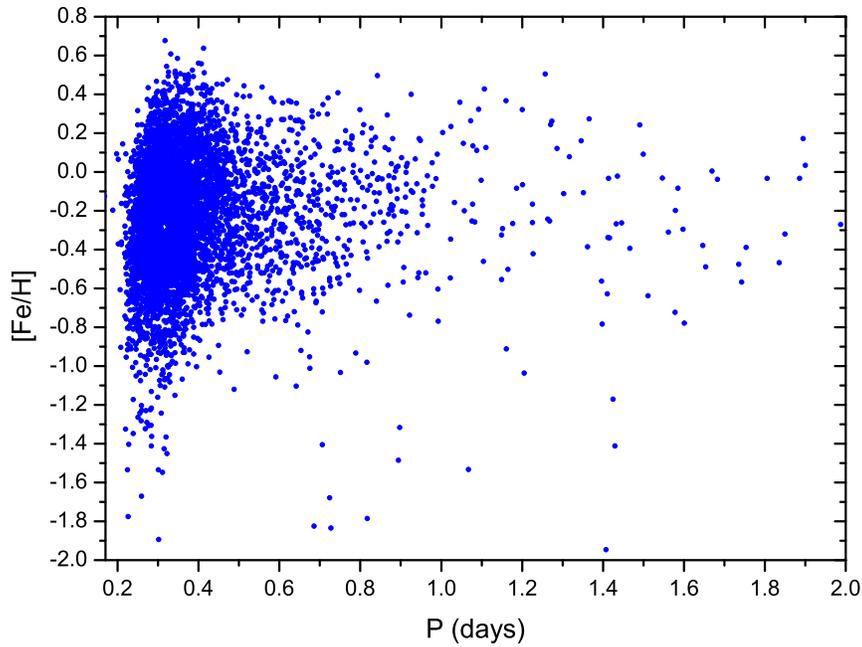}
\caption{
The same as those in Figs. 7 and 8 but for the correlation between the orbital period and the metallicity.}
\end{center}
\end{figure}

\begin{figure}
\begin{center}
\includegraphics[angle=0,scale=0.5]{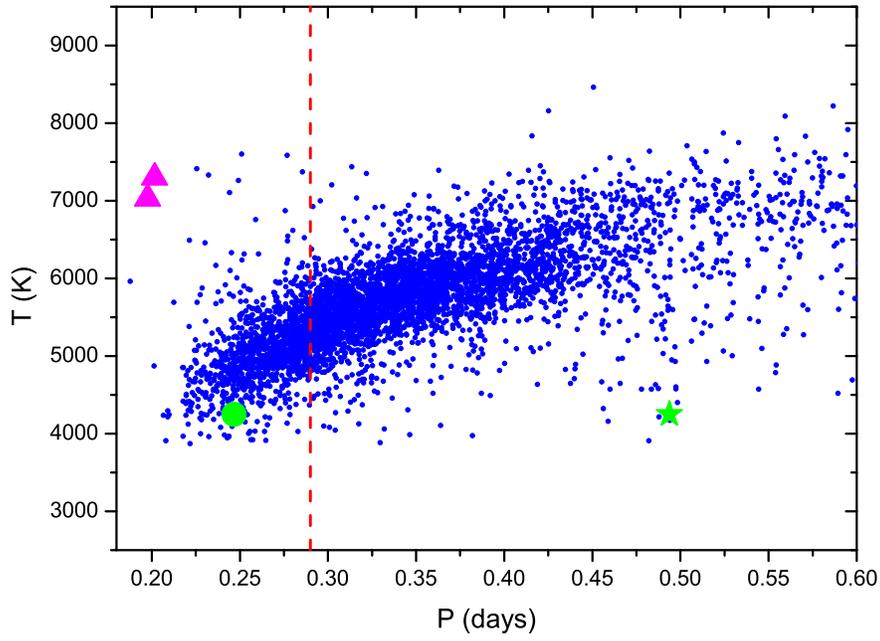}
\caption{Correlation between the orbital period and the effect temperature for EWs with orbital periods shorter than 0.6\,days. Magenta solid triangles refer to the positions of 1SWASP J193127.17+465809.1 and 1SWASP J235935.22+362001.5 that may contain bright third bodies. The green solid star represents the position of CSS\_J080814.1+184933 with a wrong period, while the green solid circle to the right position with the revised one.}
\end{center}
\end{figure}

\begin{figure}
\begin{center}
\includegraphics[angle=0,scale=.5]{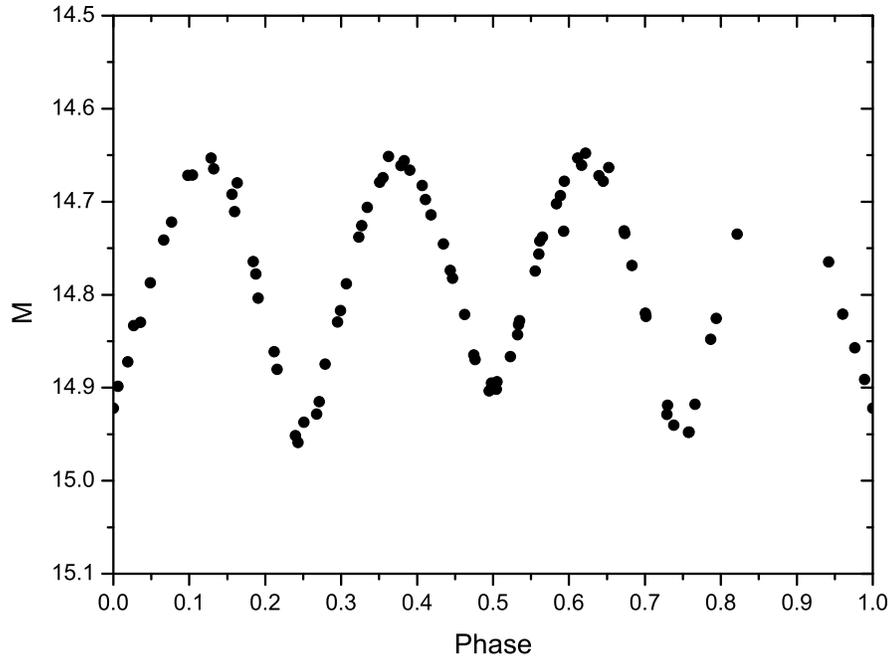}
\caption{Phased light curve of CSS\_J080814.1+184933 by using the orbital period (0.493868 days) given in VSX.}
\end{center}
\end{figure}

\begin{figure}
\begin{center}
\includegraphics[angle=0,scale=.5]{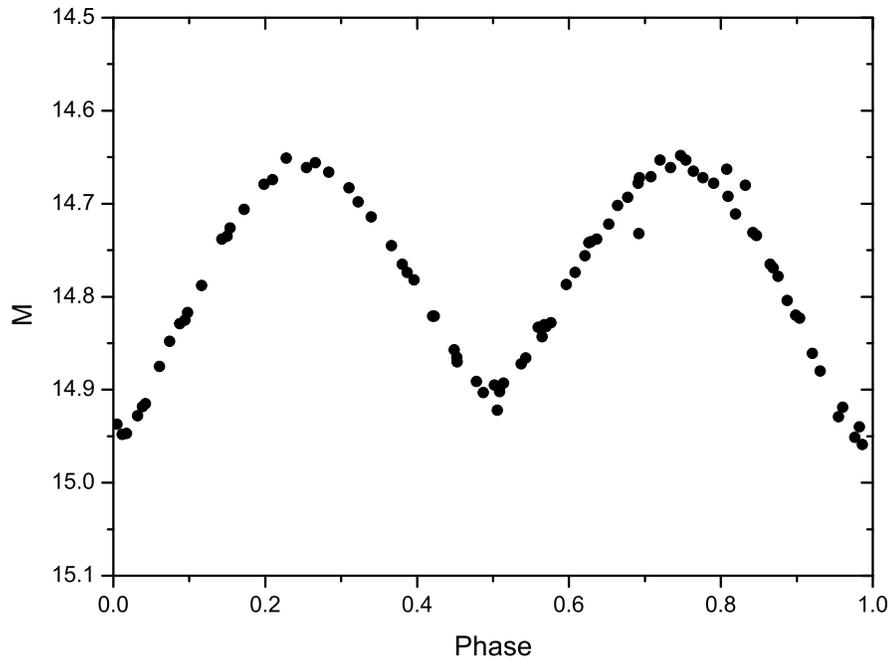}
\caption{The light curve of CSS\_J080814.1+184933. The phases were computed with the revised period 0.246746 days.}
\end{center}
\end{figure}

\begin{figure}
\begin{center}
\includegraphics[angle=0,scale=.5]{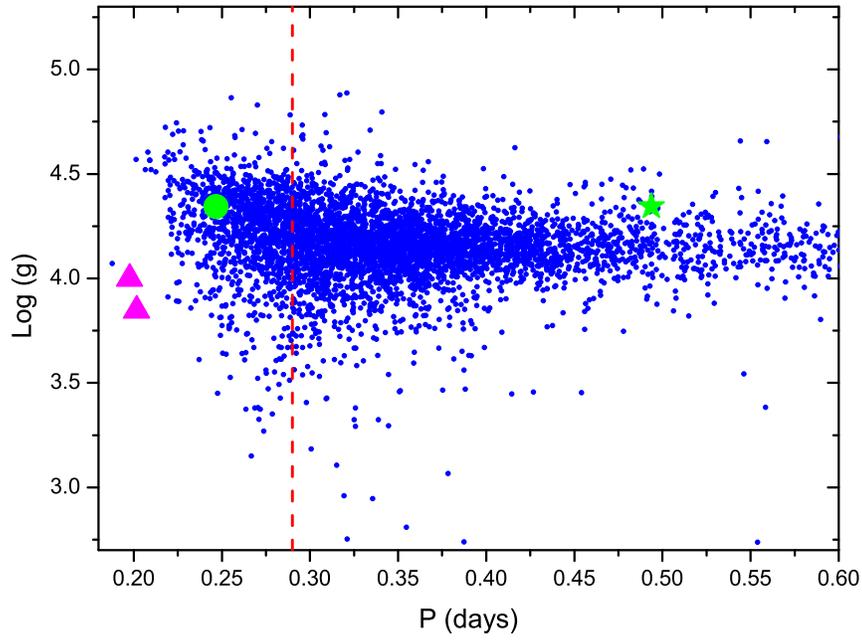}
\caption{The relation between the orbital period (P) and the gravitational acceleration Log (g) for short-period EWs. Symbols are the same as those in Fig. 10.}
\end{center}
\end{figure}

\begin{figure}
\begin{center}
\includegraphics[angle=0,scale=.5]{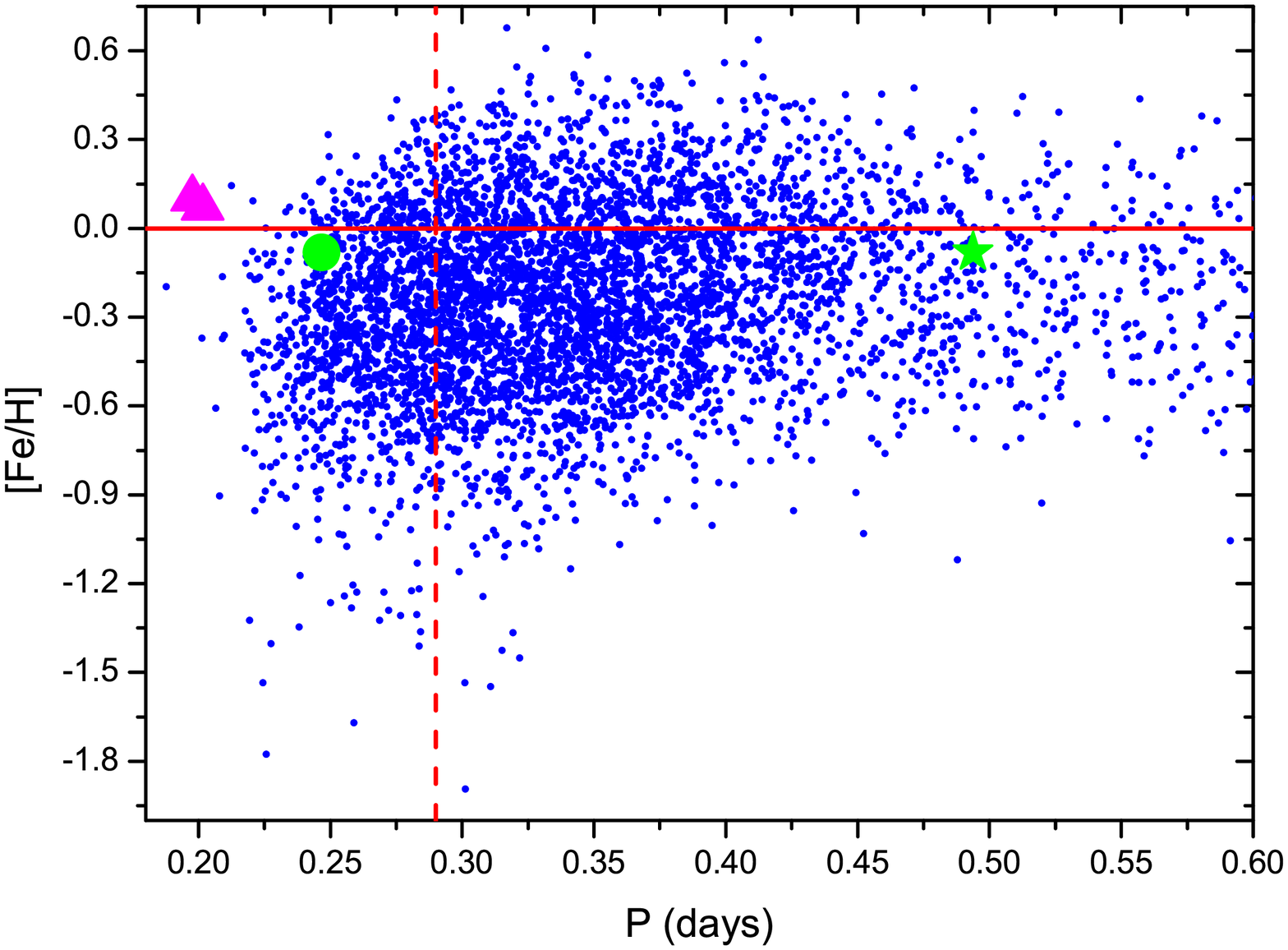}
\caption{The relation between the orbital period (P) and the metallicity [Fe/H] for short-period EWs. Symbols are the same as those in Figs. 10 and 13.}
\end{center}
\end{figure}

The correlations between the orbital period and the effect temperature $T_{eff}$, the gravitational acceleration Log (g) and the metallicity [Fe/H] are shown in Figs. 7-9 where 55 EWs with orbital period longer than 2 days are not displayed in the figures. Their orbital periods are from 2\,days to 24\,days. Six EWs are also not included in these figures because their orbital periods are unknown. Their atmosphere parameters are listed in Table 6. As shown in the three figures, the vast majority of EWs have periods in the range $0.2 < P < 0.6$ \,days. The binary components in these short-period systems are usually main-sequence stars. However, there are some EW-type contact binaries with periods over 1\,day. As plotted in Fig. 8, the gravitational acceleration (Log (g)) is weakly correlated with the orbital period. The longer is the orbital period, the lower the gravitational acceleration will be. The lower Log (g) of the long-period systems indicates that the component stars are evolved from zero-age main sequence.

\begin{table}[h]
\footnotesize
\begin{center}
\caption{LAMOST observations of EWs without orbital periods.}\label{XXXX}
\begin{tabular}{lllllllll}\hline\hline
Name & R.A. (deg) & Dec. (deg) &  Dates & Sp.  & T (K)& Log (g) &[Fe/H] & RV (Km/s) \\\hline
     V0674 Per                     & 54.115 & 36.37417 &   2014-11-13  &           K0    & 5046.24 & 4.307 & -0.162 & -39.99 \\
     V0726 Aur                     & 80.69046 & 29.11806 &   2014-11-03  &           F9    &             &             &              &                \\
     NSV 3633                      & 113.47608 & 48.00353 &   2013-11-22  &           F5    & 6327.91 & 4.174 & -0.428 & 1.75 \\
     NSV 3633                      & 113.47608 & 48.00353 &   2015-01-31  &           F5    & 6370.36 & 4.121 & -0.348 & 4.84 \\
     NSV 3633                      & 113.47608 & 48.00353 &   2016-02-19  &           F5    & 6353.8 & 4.128 & -0.362 & 14.1 \\
     NSV 5580                      & 185.6305 & 25.82833 &   2012-01-11  &           F7    & 6277.52 & 4.312 & 0.051 & 1.75 \\
     NSV 5580                      & 185.6305 & 25.82833 &   2012-02-01  &           F7    & 6230.28 & 4.341 & 0.033 & -17.33 \\
     NSV 5652                      & 187.37508 & 29.51272 &   2012-01-11  &           F7    & 6227.6 & 4.121 & 0.063 & -30.58 \\
     MG1 809127                    & 240.96542 & 2.92944 &   2016-05-10  &          A5V    &             &             &              &                \\
     Konkoly V14                   & 81.62869 & 12.95734 &   2013-10-02  &           F0    & 6857.67 & 3.474 & 0.396 & -16.07 \\
     SvkV34                        & 85.75447 & 40.95046 &   2012-02-18  &           K5    & 4520.01 & 4.382 & -0.001 & 23.93 \\
\hline\hline
\end{tabular}
\end{center}
\end{table}

Fig. 7 shows that there is a well correlation between the orbital period and the effect temperature $T_{eff}$ for short-period EWs (e.g., $P < 0.6$\,days). This is more clearly seen in Fig. 10 where only those short-period EWs are shown. The red dashed line in Fig. 10 refers to the peak value of the period distribution. The relation is similar to the period-color relation for EWs \citep[e.g.,][]{1961ROB..13....E, 1967MmRAS..70..111E, 1998AJ....116.2998R, 2012AJ....143...99T}. Both main-sequence components in short-period EWs are filling the critical Roche lobes and have a common convective envelope. It is expected that longer-period systems should have higher-mass components with higher temperatures \citep[e.g.,][]{2003MNRAS.342.1260Q}. However, this relation shows a large scatter. This may be caused by the effect of the presence of third stars. Two examples of this kind of cases are 1SWASP J193127.17+465809.1 and 1SWASP J235935.22+362001.5 \citep[e.g.,][]{2013A&A...549A..86L}. They are extremely short-period EWs with orbital periods 0.1976295 and 0.2016714\,days respectively. It is expected that they should be extremely cool binary systems. However, their spectral types determined by LAMOST are F0 with temperatures of 7027\,K and 7294\,K respectively. The possibility caused this difference is the presence of a third body with spectral type of F0. New spectroscopic and photometric data are very useful to study the two interesting systems. Their positions in Fig. 10 are shown as magenta solid triangles that deviate the general trend greatly.

The other possibility caused the large scatter in the period-temperature relation is that the periods of some binaries are wrong. One of the examples is CSS\_J080814.1+184933. Its orbital period given in VSX is 0.493868 days. The green solid star in Fig. 10 refers to its position that does not follow the general trend of the period-temperature relation. The phased light curve by using this period is shown in Fig. 11. As we seen in the figure, there are two primary minima and two secondary minima in one phased light curve. This indicates that the period is wrong. By using those new data, the period of the binary was revised as 0.246746\,days. The phased light curves with the revised period is displayed in Fig. 12 that is a typical EW-type light curve. The green solid circle in Fig. 10 represents the right position of the binary with the revised period. As shown in Fig. 10, the period-temperature relation is tight, but this relation is not linear.

The relations between the orbital period and Log (g) and [Fe/H] for short-period EWs are shown in Figs. 13 and 14. The positions of the three special EWs are also plotted in the two figures. As that in Fig. 10, the dashed lines represent the peak value of the period distribution. Fig. 13 shows that Log (g) is weakly correlated with the orbital period. The relation for short-period systems ($P < 0.29$\,days) is deeper than that for long-period ones ($P > 0.29$\,days). As displayed in Fig. 14, most of the EWs have lower metallicities (below the red solid line). The metallicity is also weakly correlated with the orbital period. For short-period EWs with period shorter than 0.25\,days, all of them have metallicities below zero.

\section{Discussions and conclusions}

Numerous EWs have been discovered through several large photometric surveys (e.g., CSS, the asteroid survey LINEAR, ASAS and NSVS). However, spectroscopic data are lack for those EWs. Among 40785 EWs listed in VSX catalogue, 7938 were observed in LAMOST spectral survey from October 24, 2011 to November 30, 2016. We catalogue those EWs and their spectral types are given. We also present stellar atmospheric parameters for 5363 of them. By analyzing 25 EWs observed five times or more by LAMOST, we show that the standard errors of the effect temperatures, the gravitational acceleration and the metallicity are usually lower than 110\,K, 0.19\,dex and 0.11\,dex, respectively. These results may indicate that the spectra of EWs have sufficient tracers necessary for unique determination of atmospheric parameters and the extracting parameters could reach the mentioned level of precision. However, to check the results obtained by LAMOST, a careful selection and a detailed spectroscopic investigation of some EWs observed by LAMOST are needed. We are observing some EWs spectroscopically and will determine their stellar atmospheric parameters and compare them with those obtained by LAMOST.

The derived effect temperatures as well as the spectral types by LAMOST are very useful during the photometric solution of their light curves. The other atmospheric parameters provide us valuable information to understand the formation and evolutionary state of EWs. We found that the peak of the period distribution is near 0.29\,days that is shorter than that given by previous investigators \citep[e.g.,][]{2006MNRAS.368.1311P}. This indicates that a large number of short-period faint EWs were discovered by recently deep photometric surveys. The distributions of the effect temperature (T), the gravitational acceleration (Log(g)), the metallicity ([Fe/H]) and the radial velocity (RV) are presented for those observed EWs. There are two peaks in the temperature distribution that correspond to p-p chain and CNO nuclear reactions in the component cores respectively. The distribution of the gravitational acceleration Log (g) indicates that the components of most EWs are main-sequence stars that is consistent with the idea that EWs are formed from detached main-sequence binaries via the combination of Case A mass transfer and angular momentum loss via magnetic braking \citep[e.g.,][]{2013ApJS..209...13Q}.

It is detected that metallicities of most sample stars (about 80.6\%) are below zero. Since stellar metallicities are weakly correlated with their ages \citep[e.g.,][]{2007ApJ...665..767R, 2009IAUS..258...23F}, this detection reveals that EW-type systems are old stellar population. This supports the assumption that EWs need a long-term pre-contact evolution with timescales from a few hundred million to a few billion years. A few percent of EWs with unusual high metallicities may be contaminated by material from the evolutionary processes of unseen neutron stars and black holes in the systems. The progenitors of those unseen degenerate objects are originally much more massive third stars in triple systems. To date, neutron stars and black holes are usually discovered in X-ray binaries through their X-ray radiation. They are formed through common envelope evolution and are influenced by the companion stars. If these unseen degenerate objects are confirmed, they will be a new population of neutron stars and black holes \citep[e.g.,][]{2008ApJ...687..466Q, 2010MmSAI..81..294Z}.

The correlation between the orbital period and the effect temperature, the gravitational acceleration and the metallicity are shown in previous sections. The scatters of those figures may be mainly caused by the presence of the third bodies. EWs have the shortest period and the lowest angular momentum among main-sequence binaries. It is assumed that they have a third body that plays an important role for their formation by removing angular momentum from the central binary \citep[e.g.,][]{2006NewA...12..117Q, 2007AJ....133..357Q, 2013ApJS..207...22Q, 2013AJ....145...39Z}. For some systems, their orbital periods are wrong that also cause the scatters in those diagrams. It is shown that the relation between the orbital period and the effect temperature is tight and non-linear.

Both the gravitational acceleration and the metallicity are weakly correlated with the orbital period. The metallicities of all short-period EWs with period shorter than 0.25\,days are below zero.
These indicate that the physical properties of EWs are mainly depending on their orbital periods. The formations and evolutionary states of EWs with different orbital period may be quite different.
This conclusion is supported by the relations between the effect temperature (T) and the gravitational acceleration Log (g) and the metallicity [Fe/H] that are shown in Figs. 15 and 16. The dashed lines in the two figures refer to the peak of the temperature distribution at 5700\,K. It is found that both the gravitational acceleration and the metallicity are weakly correlated with the effect temperature. Those extremely short-period EWs (P<0.29\,days; T<5700\,K) usually have higher gravitational acceleration Log (g) and lower metallicity [Fe/H]. They are main-sequence stars with little evolution and are older than their hotter and long-period cousins. This may suggest that they have a long timescale of pre-contact evolution and their formation and evolution are mainly driven by angular momentum loss via magnetic braking.

\begin{figure}
\begin{center}
\includegraphics[angle=0,scale=.5]{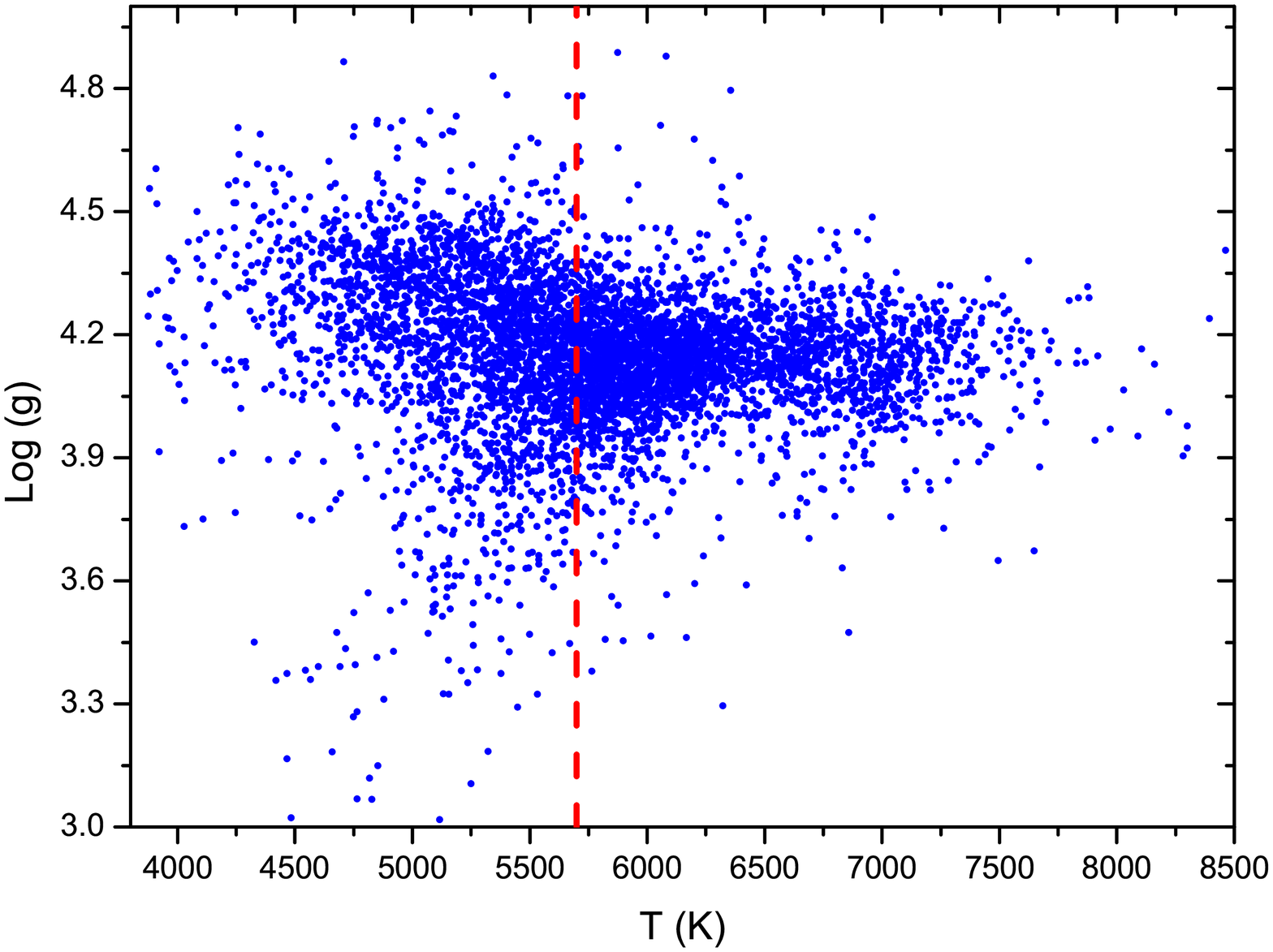}
\caption{The relation between the effect temperature (T) and the gravitational acceleration Log (g) for EWs. The red dashed line represents the peak of the temperature distribution at 5700\,K.}
\end{center}
\end{figure}

\begin{figure}
\begin{center}
\includegraphics[angle=0,scale=.5]{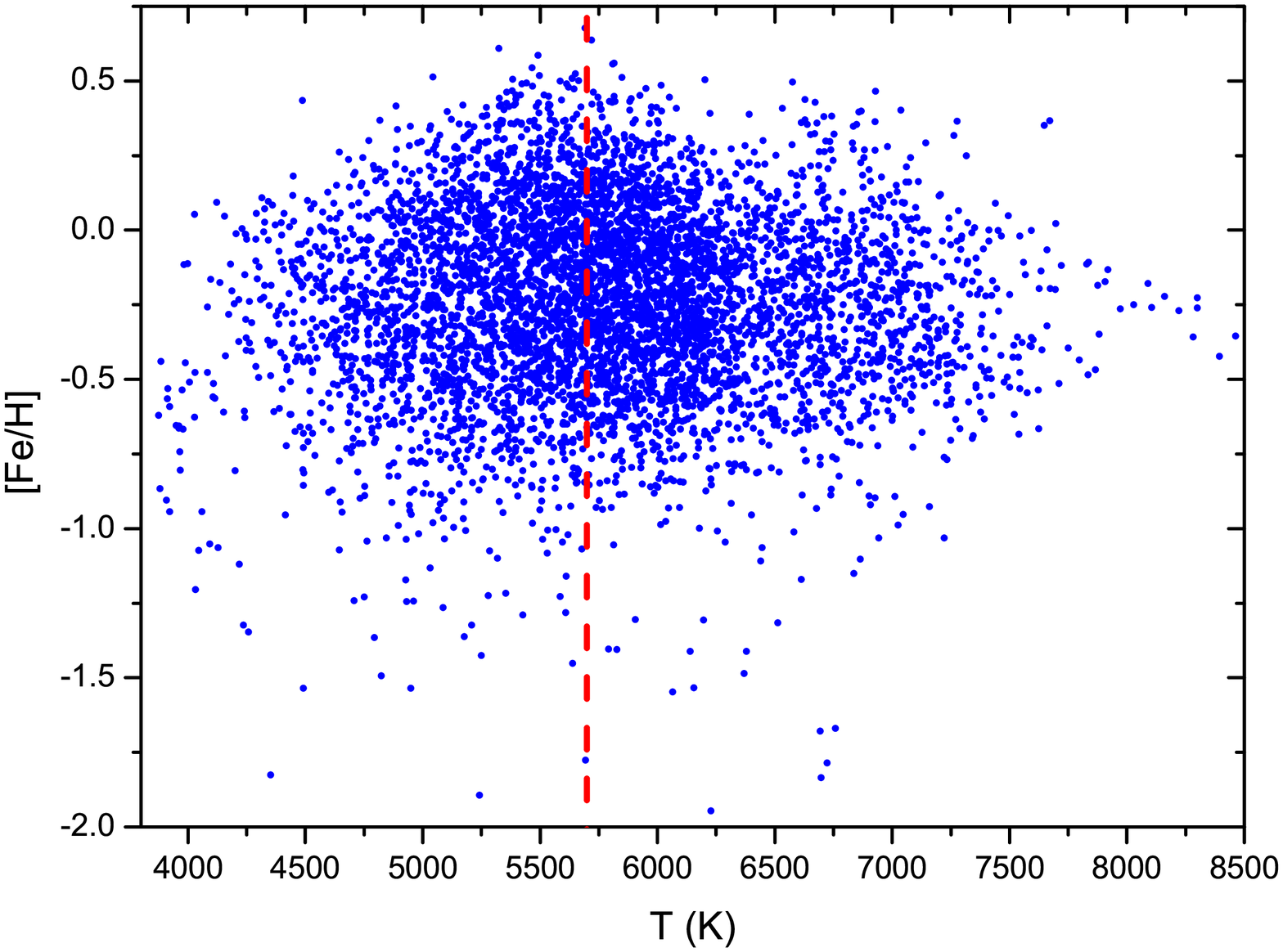}
\caption{The relation between the effect temperature (T) and the metallicity [Fe/H] for EWs. Also shown as the red dashed line is the peak of the temperature distribution.}
\end{center}
\end{figure}

\begin{acknowledgements}
This work is partly supported by Chinese Natural Science Foundation (No. 11325315). Guoshoujing Telescope (the Large Sky Area Multi-Object Fiber Spectroscopic Telescope LAMOST) is a National Major Scientific Project built by the Chinese Academy of Sciences. Funding for the project has been provided by the National Development and Reform Commission. LAMOST is operated and managed by the National Astronomical Observatories, Chinese Academy of Sciences. Spectroscopic observations used in the paper were obtained with LAMOST from October 24, 2011 to November 30, 2016.
\end{acknowledgements}

\label{lastpage}


\begin{thebibliography}{}


\bibitem[Becker et al.(2011)]{2011ApJ...731...17B} Becker, A.~C., Bochanski, J.~J., Hawley, S.~L., et al.\ 2011, \apj, 731, 17
\bibitem[Bradstreet \& Guinan(1994)]{1994ASPC...56..228B} Bradstreet, D.~H., \& Guinan, E.~F.\ 1994, Interacting Binary Stars, 56, 228
\bibitem[Cox(2000)]{2000AAQ...C} Cox, A. N., 2000, Allen's Astrophysical Quantities, 4th ed. (New York: Springer)
\bibitem[Cui et al.(2012)]{2012RAA....12.1197C} Cui, X.-Q., Zhao, Y.-H., Chu, Y.-Q., et al.\ 2012, Research in Astronomy and Astrophysics, 12, 1197
\bibitem[Drake et al.(2009)]{2009ApJ...696..870D} Drake, A.~J., Djorgovski, S.~G., Mahabal, A., et al.\ 2009, \apj, 696, 870
\bibitem[Drake et al.(2014)]{2014ApJS..213....9D} Drake, A.~J., Graham, M.~J., Djorgovski, S.~G., et al.\ 2014, \apjs, 213, 9
\bibitem[Eggen(1961)]{1961ROB..13....E} Eggen O. J., 1961, Royal Obs. Bull., 13
\bibitem[Eggen(1967)]{1967MmRAS..70..111E} Eggen, O.~J.\ 1967, \memras, 70, 111
\bibitem[Feltzing \& Bensby(2009)]{2009IAUS..258...23F} Feltzing, S., \& Bensby, T.\ 2009, The Ages of Stars, 258, 23
\bibitem[Gao et al.(2015)]{2015RAA....15.2204G} Gao, H., Zhang, H.-W., Xiang, M.-S., et al.\ 2015, Research in Astronomy and Astrophysics, 15, 2204
\bibitem[Guinan \& Bradstreet(1988)]{1988FE....345G} Guinan E. F. \& Bradstreet D. H., 1988, in: Dupree A. K., Lago M. T. V. T. (eds) Formation and Evolution of Low Mass Star. Kluwer Acad. Publ., P345
\bibitem[Kaluzny \& Rucinski(1993)]{1993BS....345G} Kaluzny, J. \& Rucinski, S. M., 1993, in ASP Conf. Ser. 53, Blue Stragglers, ed. R. A. Saffer (San Francisco, CA: ASP), 164
\bibitem[Koleva et al.(2009)]{2009A&A...501.1269K} Koleva, M., Prugniel, P., Bouchard, A., \& Wu, Y.\ 2009, \aap, 501, 1269
\bibitem[Lohr et al.(2013)]{2013A&A...549A..86L} Lohr, M.~E., Norton, A.~J., Kolb, U.~C., et al.\ 2013, \aap, 549, A86
\bibitem[Luo et al.(2012)]{2012RAA....12.1243L} Luo, A.-L., Zhang, H.-T., Zhao, Y.-H., et al.\ 2012, Research in Astronomy and Astrophysics, 12, 1243
\bibitem[Luo et al.(2015)]{2015RAA....15.1095L} Luo, A.-L., Zhao, Y.-H., Zhao, G., et al.\ 2015, Research in Astronomy and Astrophysics, 15, 1095
\bibitem[Paczy{\'n}ski et al.(2006)]{2006MNRAS.368.1311P} Paczy{\'n}ski, B., Szczygie{\l}, D.~M., Pilecki, B., \& Pojma{\'n}ski, G.\ 2006, \mnras, 368, 1311
\bibitem[Palaversa et al.(2013)]{2013AJ....146..101P} Palaversa, L., Ivezi{\'c}, {\v Z}., Eyer, L., et al.\ 2013, \aj, 146, 101
\bibitem[Pojmanski(1997)]{1997AcA....47..467P} Pojmanski, G.\ 1997, \actaa, 47, 467
\bibitem[Pojmanski et al.(2005)]{2005AcA....55..275P} Pojmanski, G., Pilecki, B., \& Szczygiel, D.\ 2005, \actaa, 55, 275
\bibitem[Prugniel \& Soubiran(2001)]{2001A&A...369.1048P} Prugniel, P., \& Soubiran, C.\ 2001, \aap, 369, 1048
\bibitem[Prugniel et al.(2007)]{2007astro.ph..3658P} Prugniel, P., Soubiran, C., Koleva, M., \& Le Borgne, D.\ 2007, arXiv:astro-ph/0703658
\bibitem[Qian(2003)]{2003MNRAS.342.1260Q} Qian, S.\ 2003, \mnras, 342, 1260
\bibitem[Qian et al.(2008)]{2008ApJ...687..466Q} Qian, S.-B., Liao, W.-P., \& Fern{\'a}ndez Laj{\'u}s, E.\ 2008, \apj, 687, 466
\bibitem[Qian et al.(2006)]{2006NewA...12..117Q} Qian, S.-B., Liu, L., \& Kreiner, J.~M.\ 2006, \na, 12, 117
\bibitem[Qian et al.(2013a)]{2013ApJS..209...13Q} Qian, S.-B., Liu, N.-P., Li, K., et al.\ 2013a, \apjs, 209, 13
\bibitem[Qian et al.(2014)]{2014ApJS..212....4Q} Qian, S.-B., Wang, J.-J., Zhu, L.-Y., et al.\ 2014, \apjs, 212, 4
\bibitem[Qian et al.(2007)]{2007AJ....133..357Q} Qian, S.-B., Xiang, F.-Y., Zhu, L.-Y., et al.\ 2007, \aj, 133, 357
\bibitem[Qian et al.(2013b)]{2013ApJS..207...22Q} Qian, S.-B., Zhang, J., Wang, J.-J., et al.\ 2013b, \apjs, 207, 22
\bibitem[Reid et al.(2007)]{2007ApJ...665..767R} Reid, I.~N., Turner, E.~L., Turnbull, M.~C., Mountain, M., \& Valenti, J.~A.\ 2007, \apj, 665, 767
\bibitem[Rucinski(1998)]{1998AJ....116.2998R} Rucinski, S.~M.\ 1998, \aj, 116, 2998
\bibitem[Rucinski et al.(2001)]{2001AJ....122.1974R} Rucinski, S.~M., Lu, W., Mochnacki, S.~W., Og{\l}oza, W., \& Stachowski, G.\ 2001, \aj, 122, 1974
\bibitem[Rucinski et al.(2013)]{2013AJ....146...70R} Rucinski, S.~M., Pribulla, T., \& Budaj, J.\ 2013, \aj, 146, 70
\bibitem[Samus et al.(2017)]{2017ARep...61...80S} Samus', N.~N., Kazarovets, E.~V., Durlevich, O.~V., Kireeva, N.~N., \& Pastukhova, E.~N.\ 2017, Astronomy Reports, 60, No. 1
\bibitem[Terrell et al.(2012)]{2012AJ....143...99T} Terrell, D., Gross, J., \& Cooney, W.~R.\ 2012, \aj, 143, 99
\bibitem[Wang et al.(1996)]{1996ApOpt..35.5155W} Wang, S.-G., Su, D.-Q., Chu, Y.-Q., Cui, X., \& Wang, Y.-N.\ 1996, \ao, 35, 5155
\bibitem[Watson(2006)]{2006SASS...25...47W} Watson, C.~L.\ 2006, Society for Astronomical Sciences Annual Symposium, 25, 47
\bibitem[Wo{\'z}niak et al.(2004)]{2004AJ....127.2436W} Wo{\'z}niak, P.~R., Vestrand, W.~T., Akerlof, C.~W., et al.\ 2004, \aj, 127, 2436
\bibitem[Wu et al.(2011a)]{2011A&A...525A..71W} Wu, Y., Singh, H.~P., Prugniel, P., Gupta, R., \& Koleva, M.\ 2011a, \aap, 525, A71
\bibitem[Wu et al.(2011b)]{2011RAA....11..924W} Wu, Y., Luo, A.-L., Li, H.-N., et al.\ 2011b, Research in Astronomy and Astrophysics, 11, 924
\bibitem[Wu et al.(2014)]{2014IAUS..306..340W} Wu, Y., Du, B., Luo, A., Zhao, Y., \& Yuan, H.\ 2014, Statistical Challenges in 21st Century Cosmology, 306, 340
\bibitem[Zhao et al.(2012)]{2012RAA....12..723Z} Zhao, G., Zhao, Y.-H., Chu, Y.-Q., Jing, Y.-P., \& Deng, L.-C.\ 2012, Research in Astronomy and Astrophysics, 12, 723
\bibitem[Zhu et al.(2013b)]{2013AJ....145...39Z} Zhu, L.~Y., Qian, S.~B., Liu, N.~P., Liu, L., \& Jiang, L.~Q.\ 2013b, \aj, 145, 39
\bibitem[Zhu et al.(2013a)]{2013AJ....146...28Z} Zhu, L.-Y., Qian, S.-B., Zhou, X., et al.\ 2013a, \aj, 146, 28
\bibitem[Ziolkowski(2010)]{2010MmSAI..81..294Z} Ziolkowski, J.\ 2010, \memsai, 81, 294


\end{thebibliography}
\end{document}